\documentclass[floatfix,twocolumn,showpacs,preprintnumbers,amsmath,amssymb,pra,superscriptaddress,longbibliography]{revtex4-1}
\usepackage{color}
\usepackage[usenames,dvipsnames,svgnames,table]{xcolor}
\usepackage[colorlinks=true,linkcolor=blue,urlcolor=blue,citecolor=blue]{hyperref}
\usepackage{mathtools}
\usepackage{graphicx}
\usepackage{dcolumn}
\usepackage{array}
\usepackage{lipsum}
\usepackage{bm}
\usepackage{subfigure}
\usepackage{amssymb}
\usepackage{multirow}
\usepackage{tabularx}
\usepackage{amsmath}
\usepackage{braket}
\graphicspath{{plots/}}
 \usepackage{lipsum}
\usepackage{mathrsfs}

\begin{document}
\title{
Electronic Density Response of Warm Dense Hydrogen:\\ \emph{Ab initio} Path Integral Monte Carlo Simulations
}

\author{Maximilian B\"ohme}
\affiliation{Center for Advanced Systems Understanding (CASUS), D-02826 G\"orlitz, Germany}
\affiliation{Helmholtz-Zentrum Dresden-Rossendorf (HZDR), D-01328 Dresden, Germany}

\affiliation{Technische  Universit\"at  Dresden,  D-01062  Dresden,  Germany}

\author{Zhandos A.~Moldabekov}

\affiliation{Center for Advanced Systems Understanding (CASUS), D-02826 G\"orlitz, Germany}

\affiliation{Helmholtz-Zentrum Dresden-Rossendorf (HZDR), D-01328 Dresden, Germany}

\author{Jan Vorberger}
\affiliation{Helmholtz-Zentrum Dresden-Rossendorf (HZDR), D-01328 Dresden, Germany}

\author{Tobias Dornheim}
\email{t.dornheim@hzdr.de}

\affiliation{Center for Advanced Systems Understanding (CASUS), D-02826 G\"orlitz, Germany}

\affiliation{Helmholtz-Zentrum Dresden-Rossendorf (HZDR), D-01328 Dresden, Germany}

\begin{abstract}
The properties of hydrogen under extreme conditions are important for many applications, including inertial confinement fusion and astrophysical models. A key quantity is given by the electronic density response to an external perturbation, which is probed in X-ray Thomson scattering (XRTS) experiments---the state of the art diagnostics from which system parameters like the free electron density $n_e$, the electronic temperature $T_e$, and the charge state $Z$ can be inferred. In this work, we present highly accurate path integral Monte Carlo (PIMC) results for the electronic density response of hydrogen. We obtain the exchange--correlation (XC) kernel $K_{xc}$, which is of central relevance for many applications, such as time-dependent density functional theory (TD-DFT). This gives us a first unbiased look into the electronic density response of hydrogen in the warm-dense matter regime, thereby opening up a gamut of avenues for future research.
\end{abstract}

\maketitle

Matter at extreme densities ($n\sim10^{21-27}/$cm$^{-3}$) and temperatures ($T\sim10^{4-8}$K) is ubiquitous throughout our universe~\cite{fortov_review}. This so-called \emph{warm dense matter}~\cite{wdm_book} (WDM) naturally occurs in a number of astrophysical objects such as giant planet interiors~\cite{Benuzzi_Mounaix_2014,Militzer_2008}, brown dwarfs~\cite{saumon1}, and the outer layers of neutron stars~\cite{neutron_star_envelopes,Haensel}. In addition, WDM is highly important for technological applications such as the discovery of novel materials~\cite{Kraus2017,Lazicki2021} and hot-electron chemistry~\cite{Brongersma2015}. Moreover, a hydrogen fuel capsule has to traverse the WDM regime~\cite{hu_ICF} on its compression path towards inertial confinement fusion (ICF)~\cite{Betti2016}.
Consequently, WDM is a highly active research area 
and a strongly increasing number of experiments~\cite{falk_wdm} 
are performed at large research facilities such as the National Ignition Facility (NIF) at the Lawrence Livermore National Laboratory~\cite{Moses_NIF} or the  European X-FEL in Hamburg, Germany~\cite{Tschentscher_2017}. 
These developments have led to a number of spectacular recent discoveries,
including the high-precision measurement of the stopping power in WDM~\cite{Cayzac2017}, the use of record peak-brightness free-electron lasers to study shock compressed aluminium~\cite{Fletcher2015}, and the possible observation of the molecular-to-metallic transition in hydrogen at extreme pressure~\cite{Knudson_Science_2015,Dias_Silvera_Science_2017}.

At the same time, we stress that the interpretation of such experiments decisively depends on a rigorous theoretical modelling. In fact, even basic system parameters like the density or electronic temperature are generally unknown and have to be inferred. In this regard, the X-ray Thomson scattering (XRTS) technique~\cite{siegfried_review,kraus_xrts} has emerged as the most promising method of diagnostics of WDM experiments.

Unfortunately, the theoretical description of WDM constitutes a most formidable challenge due to the highly nontrivial interplay of a number of physical effects~\cite{wdm_book,new_POP,review}. WDM states are correlated due to the Coulomb interaction, 
the electrons (and sometimes the nuclei) are partially quantum degenerate, 
 and WDM is a highly excited state, which rules out the well-stocked arsenal of ground-state quantum many-body methods~\cite{Jones_RevModPhys_2015,Foulkes_RMP_2001}.
This intricacy is usually expressed in terms of two characteristic parameters, that are both of the order of unity~\cite{Ott2018}: a) the density parameter $r_s=\overline{r}/a_\textnormal{B}$, where $\overline{r}$ and $a_\textnormal{B}$ are the average electronic distance and first Bohr radius, and b) the degeneracy temperature $\theta=k_\textnormal{B}T/E_\textnormal{F}$, with $E_\textnormal{F}$ being the Fermi energy~\cite{quantum_theory}.

\begin{figure*}
\centering\hspace*{-0.034\textwidth}
\includegraphics[width=.37\linewidth]{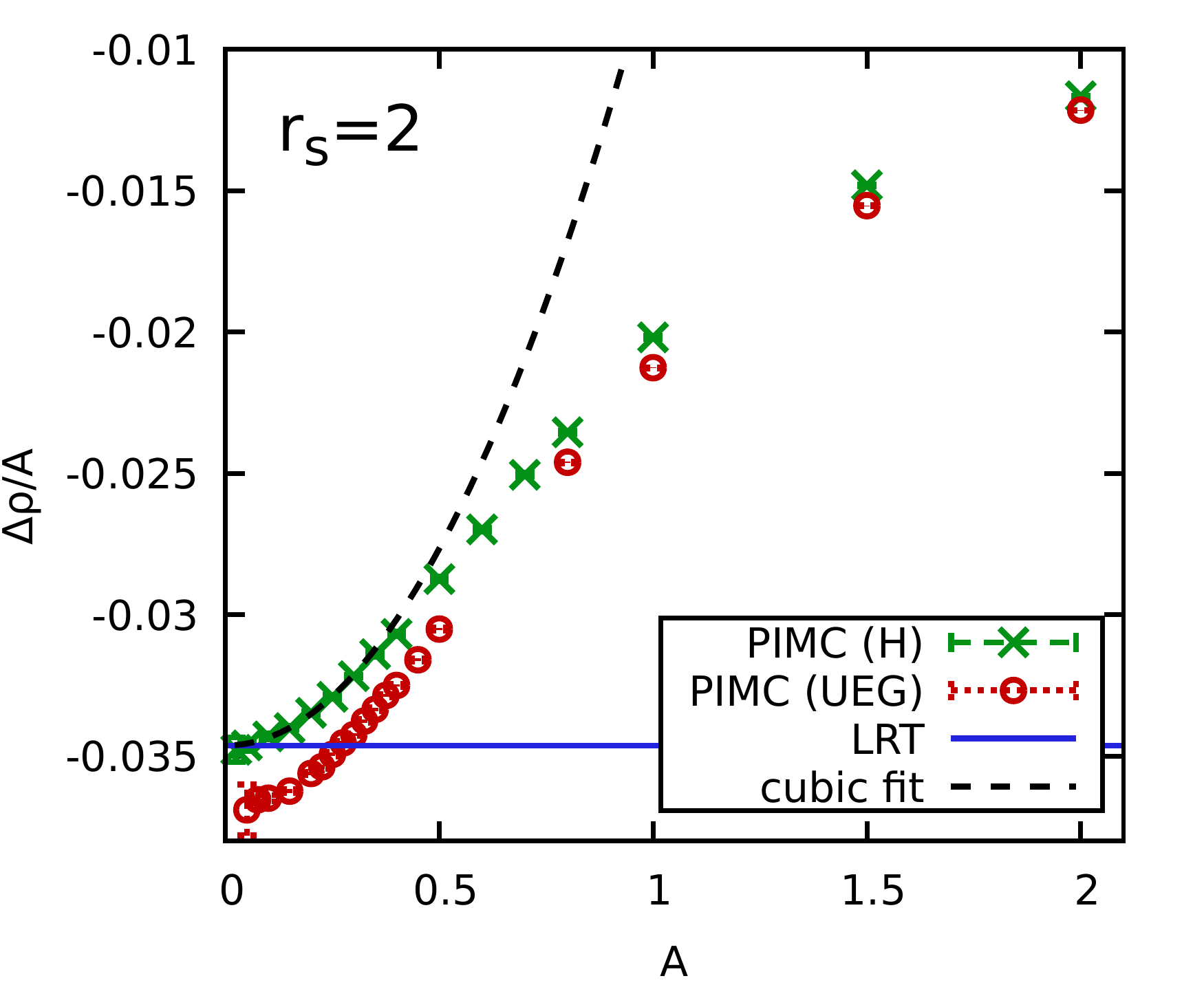}\hspace*{-0.0225\textwidth}\includegraphics[width=.37\linewidth]{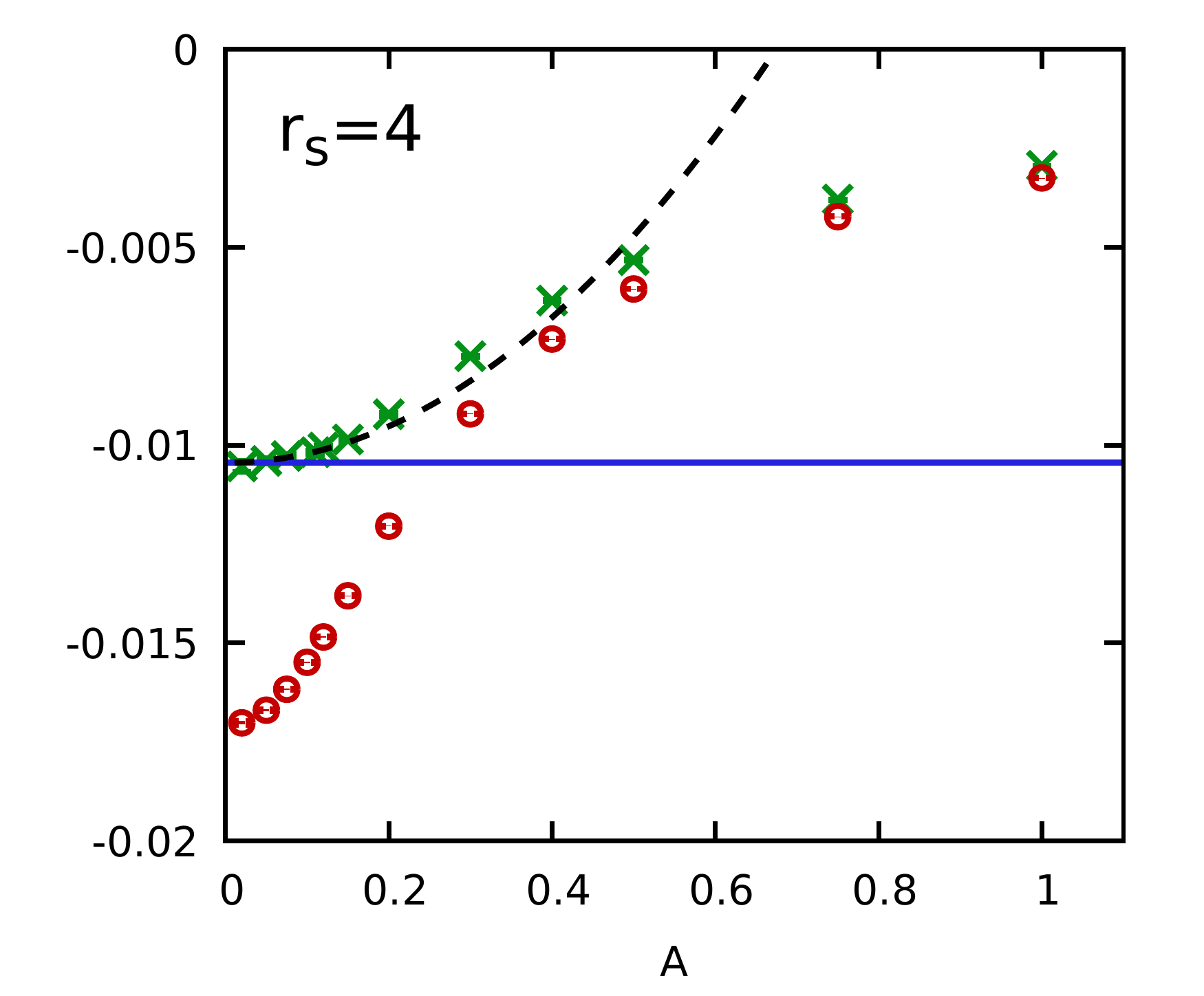}\hspace*{-0.0225\textwidth}\includegraphics[width=.37\linewidth]{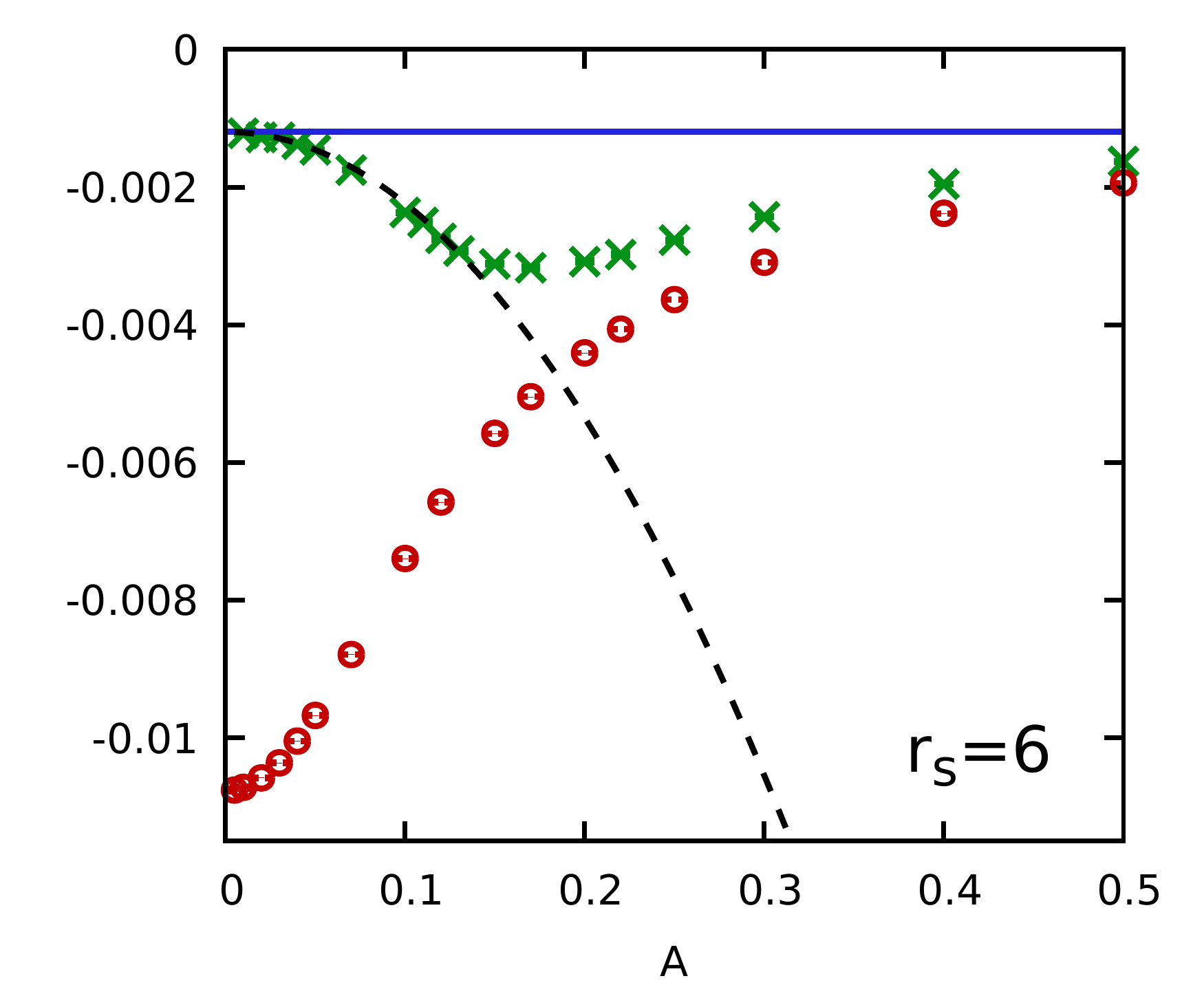}
\caption{Perturbation strength dependence of the electronic density response of hydrogen at $\mathbf{q}=2\pi L^{-1}(0,0,2)^T$ ($q\approx1.5q_\textnormal{F}$) at the electronic Fermi temperature ($\theta=1$) for $r_s=2$ (left), $r_s=4$ (center), and $r_s=6$ (right). The green crosses show our new PIMC data for hydrogen, and the red circles corresponding results for a uniform electron gas partly taken from Ref.~\cite{Dornheim_PRL_2020}. The dashed black line shows a cubic fit to the PIMC data, and the solid blue horizontal lines the linear-response limit.}
\label{fig:LRT_Limes}
\end{figure*}

In this situation, thermal density functional theory (DFT)~\cite{Mermin_DFT_1965} has emerged as the work horse in WDM theory~\cite{wdm_book}, as it often combines a manageable computation cost with a reasonable degree of accuracy. Yet, the results can substantially depend on the employed exchange--correlation (XC) functional~\cite{Burke_Perspective_JCP_2012}, which cannot be obtained within DFT itself, and has to be supplied as a semi-empirical \emph{a-priori} input; see Ref.~\cite{Pierleoni_PNAS_2016} for a critical discussion focusing on hydrogen.


An additional challenge pertaining to the interpretation of XRTS experiments is the wave-number dependence of the measured signal. The \emph{ab initio} estimation~\cite{dornheim_dynamic} of the corresponding dynamic structure factor $S(\mathbf{q},\omega)$ (beyond model assumptions such as the Chihara decomposition~\cite{Baczewski_PRL_2016,Chihara_1987}) requires one to carry out time-dependent DFT (TD-DFT) simulations~\cite{ullrich2012time}, which need as input the frequency- and wave vector-dependent XC-kernel $K_\textnormal{xc}(\mathbf{q},\omega)$. Yet, little is known about the actual $K_\textnormal{xc}(\mathbf{q},\omega)$ of a real WDM system, and previous studies~\cite{Baczewski_PRL_2016,Ramakrishna_PRB_2021} have been exclusively based on drastic simplifications such as the adiabatic local density approximation (ALDA). This is highly problematic, since these approximate kernels have been designed for the application at $T=0$ and, in the case of ALDA, are valid only for small wave numbers $q=|\mathbf{q}|$ and break down for strong degrees of inhomogeneity~\cite{Moldabekov_JCP_2021}.

In the present work, we aim to overcome these limitations by performing \emph{ab initio} path integral Monte Carlo (PIMC) calculations~\cite{cep,Dornheim_permutation_cycles} for the electronic density response of warm dense hydrogen. This allows us to thoroughly address a number of fundamental questions, including i) the quantification of the impact of electron--ion correlations on electronic properties, ii) the assessment of the validity range of linear response theory and the importance of nonlinear effects~\cite{Dornheim_PRL_2020,Dornheim_PRR_2021,Dornheim_JPSJ_2021}, and iii) to check widespread model assumptions about the decomposition into \emph{bound} and \emph{free} electrons~\cite{kraus_xrts}. Most notably, we present the first, highly accurate results for the XC-kernel of a realistic WDM system in the static limit (i.e., $\omega\to0$). 
Our theoretical predictions are directly useful for upcoming experiments with hydrogen, for example at NIF or the European X-FEL.

\textbf{Results:} We have carried out thermal DFT-MD simulations to obtain a set of ionic configurations, and use the PIMC method to obtain exact solutions to the electronic problem in the external ionic potential. This allows us to directly compare our PIMC results to DFT, and use the former as input for the latter. We note that we impose no nodal restrictions on the thermal density matrix~\cite{Ceperley1991}. Therefore, our PIMC simulations are without approximation, but computationally expensive due to the notorious fermion sign problem~\cite{troyer,dornheim_sign_problem}. We estimate the full computation cost of the present study to be of the order of $\mathcal{O}\left(10^7\right)$ CPUh.
To compute the electronic density response, we apply an external harmonic perturbation~\cite{moroni,bowen2,dornheim_pre,Dornheim_PRL_2020} of wave vector $\mathbf{q}$ and perturbation amplitude $A$, which leads to the full electronic Hamiltonian (we assume Hartree atomic units throughout this work)
\begin{eqnarray}\label{eq:Hamiltonian}
\hat{H} = -\frac{1}{2}\sum_{l=1}^N \nabla^2_l + \hat{W}_\textnormal{ee} + \hat{V}_\textnormal{ei} + 2 A \sum_{l=1}^N \textnormal{cos}\left(\mathbf{q}\cdot\hat{\mathbf{r}}_l\right)\ .
\end{eqnarray}
Here $N=N^\uparrow+N^\downarrow$ is the total number of electrons, and we restrict ourselves to the unpolarized case of $N^\uparrow=N^\downarrow=N/2$. The operators $\hat{W}_\textnormal{ee}$ and $\hat{V}_\textnormal{ei}$ denote the interaction between the electrons and the external potential due to the fixed ions, respectively.
We use the PIMC method to compute the expectation value of the electronic density in Fourier space
$\rho(\mathbf{q},A) = \left<\hat{\rho}_\mathbf{q} \right>_{\mathbf{q},A}$,
and the sought-after density response is simply given by $\Delta\rho(\mathbf{q},A)=\rho(\mathbf{q},A)-\rho(\mathbf{q},0)$.

The results for the density response of warm dense hydrogen are shown in Fig.~\ref{fig:LRT_Limes} for $r_s=2$ (left), $r_s=4$ (center), and $r_s=6$ (right) at the electronic Fermi temperature, i.e., $\theta=1$. Specifically, we show $\Delta\rho/A$, and the green crosses (red circles) depict our new PIMC results for hydrogen (for a uniform electron gas (UEG)~\cite{Dornheim_PRL_2020}). The main qualitative trends are as follows. For the metallic density of $r_s=2$, the electrons are only weakly localized around the ions, and the density response both qualitatively and quantitatively resembles the behaviour of a UEG at the same conditions. The central panel corresponds to a lower electronic density as it is realised experimentally for example within hydrogen jets~\cite{Zastrau}. In this case, a sizeable fraction of the electrons is assumed to be involved in \emph{bound states} with the protons. Consequently, we observe a starkly reduced density response compared to the UEG, as only the \emph{unbound} (free) electrons react to the external perturbation; see the discussion of Fig.~\ref{fig:Response} below. Furthermore, we observe that the data sets for hydrogen and the UEG converge in the limit of large $A$, since the external perturbation will eventually predominate over the ionic potential. Lastly, the right panel shows results for $r_s=6$. While being hard to probe in current experimental set-ups, these conditions are interesting as a challenging benchmark for theoretical methods due to the pronounced impact of electronic XC-effects~\cite{low_density1,low_density2}. In this case, we find that the electronic density response is reduced by an order of magnitude compared to the UEG at small $A$, due to the high probability of bound states.
An additional advantage of the present approach is its capability to study nonlinear effects and, in this way, to unambiguously assess the validity range of linear response theory. Overall, we find similar trends for hydrogen as in previous investigations of the UEG~\cite{Dornheim_PRL_2020,Dornheim_PRR_2021}, and nonlinear effects are even increased in hydrogen compared to the UEG at $r_s=6$. 


\begin{figure}
\centering
\includegraphics[width=.89\linewidth]{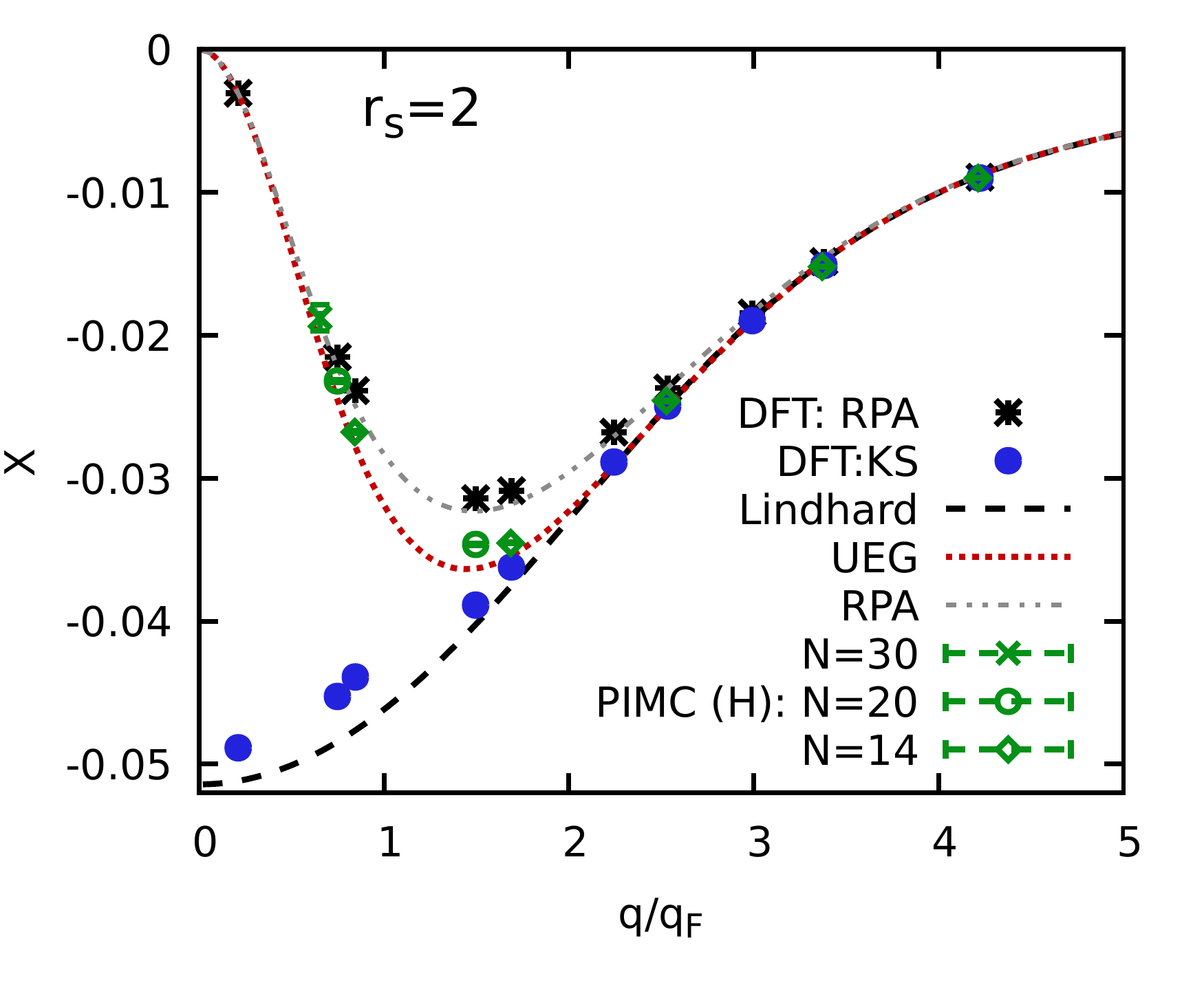}\\\vspace*{-1.17cm}\includegraphics[width=.89\linewidth]{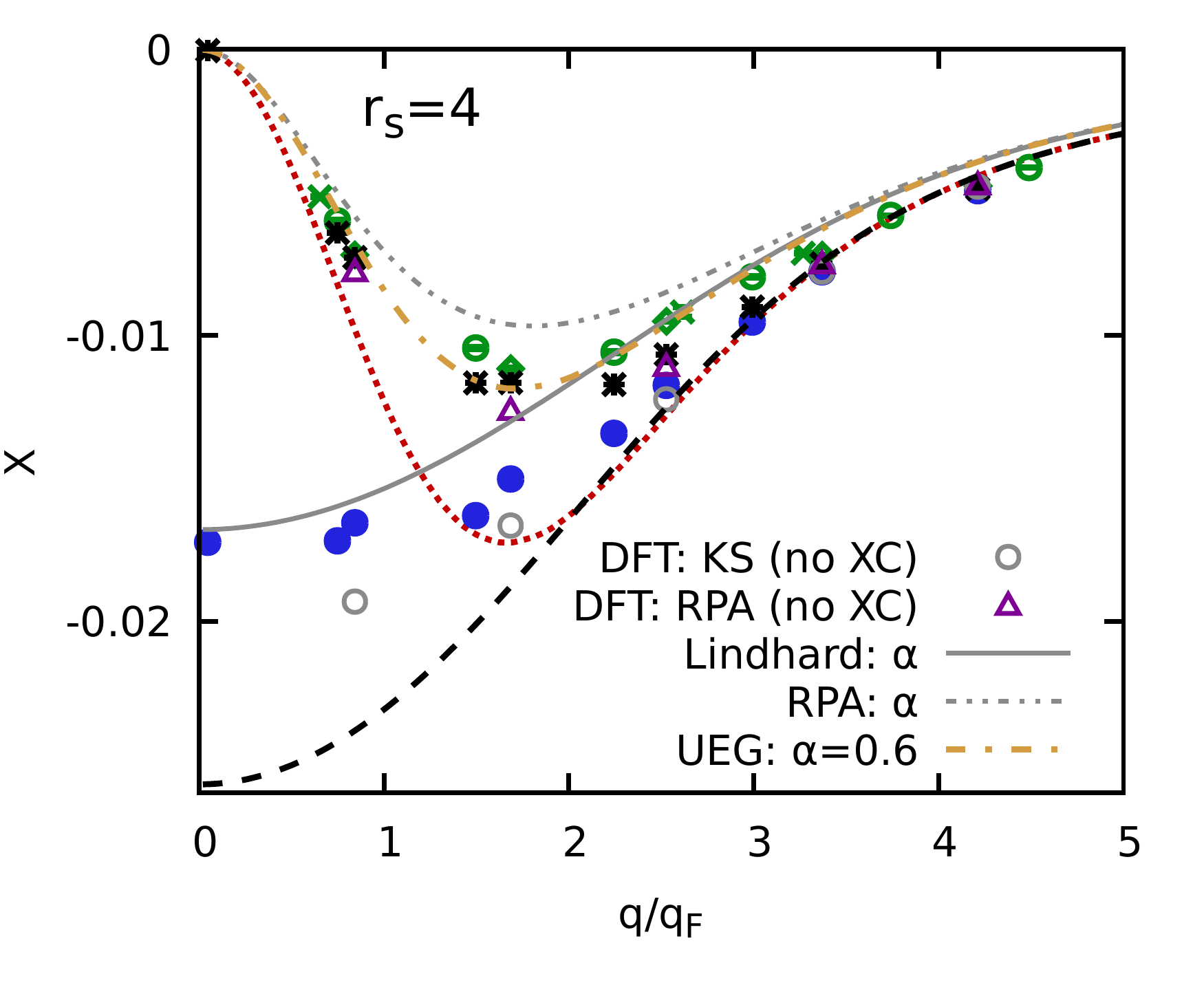}
\caption{Electronic density response of warm dense hydrogen at $\theta=1$ with $r_s=2$ (top) and $r_s=4$ (bottom). Green symbols: new PIMC results for different $N$; blue dots: Kohn-Sham response function (within LDA) $\chi_0(\mathbf{q})$; grey stars: DFT-response within RPA; dotted red: UEG model~\cite{dornheim_ML}; dashed black: Lindhard function. Also shown are results for the Kohn-Sham response function without XC-effects (grey circles), the corresponding RPA (purple triangles), the UEG model assuming a free-electron fraction of $\alpha=0.6$ (dash-dotted yellow), and the corresponding Lindhard function (solid grey) for $r_s=4$.}
\label{fig:Response}
\end{figure}

The central task of the present work is the estimation of the linear electronic density response function. The density response at the first harmonic of the original perturbation (i.e., at the same $\mathbf{q}$) can be expanded in powers of the perturbation amplitude~\cite{Dornheim_PRR_2021} as $\Delta\rho(\mathbf{q},A)=\chi(\mathbf{q})A+\chi^{\textnormal{cubic}}(\mathbf{q})A^3$, and the dashed black curves depict corresponding fits~\cite{it_should_be_noted} to the PIMC data for sufficiently small $A$. The horizontal blue lines depict the linear coefficient $\chi(\mathbf{q}):=\chi(\mathbf{q},0)$, i.e., the static limit of the linear density response function~\cite{kugler1}
\begin{eqnarray}\label{eq:chi}
\chi(\mathbf{q},\omega) = \frac{\chi_0(\mathbf{q},\omega)}{1-\frac{4\pi}{q^2}\left[1-G(\mathbf{q},\omega)\right]\chi_0(\mathbf{q},\omega)}\ .
\end{eqnarray}
Here $\chi_0(\mathbf{q},\omega)$ can be either the Kohn-Sham response function~\cite{supplement}, or the Lindhard function~\cite{quantum_theory} in the case of a uniform system. In addition, $G(\mathbf{q},\omega)=-\frac{q^2}{4\pi}K_\textnormal{xc}(\mathbf{q},\omega)$ denotes the local field correction (LFC), which contains the full wave-vector and frequency-resolved information about electronic XC-effects, and which is equivalent to the XC-kernel $K_\textnormal{xc}$ usually employed in the context of TD-DFT~\cite{ullrich2012time}. Evidently, both $G(\mathbf{q},\omega)$ and $K_\textnormal{xc}(\mathbf{q},\omega)$ depend on the particular choice of $\chi_0(\mathbf{q},\omega)$ and, in the context of DFT, on the employed XC-functional. 

The relevant linear electronic density response is shown in Fig.~\ref{fig:Response}. 
The green symbols show our new PIMC results for different $N$ that have been obtained by following the fitting procedure of $\Delta \rho(\mathbf{q},A)$ outlined above for different wave vectors $\mathbf{q}$. Evidently, no systematic dependence on the system size can be resolved, which is consistent to previous findings for the UEG~\cite{dornheim_ML,dynamic_folgepaper,dornheim_electron_liquid}. The density response of the UEG has been included as dotted red curves and is in good agreement to the green symbols at $r_s=2$; we only observe small deviations around $q\sim1.5q_\textnormal{F}$. In stark contrast, there appear pronounced differences between the UEG and hydrogen (exceeding $30\%$) at $r_s=4$, which is a direct consequence of the increased location of the electrons around the ions. It is often assumed that the total number of electrons can be decomposed into a \emph{bound} and into an effectively \emph{free} fraction; we denote the latter as $\alpha=N_\textnormal{free}/N$ in this work. This leads to the modified parameters $r_s(\alpha)$ and $\theta(\alpha)$. Empirically, we find that the choice of $\alpha=0.6$ leads to a good qualitative agreement between the UEG model and our PIMC results for hydrogen at $r_s=4$ for small to intermediate wave numbers, see the dash-dotted yellow curve in the bottom panel of Fig.~\ref{fig:Response}. This is close to the result of $\alpha\approx0.54$ by Militzer and Ceperley~\cite{Militzer_PRE_2001} based on a cluster analysis in real space. Interestingly, the agreement between the PIMC data for hydrogen and the effective free electron model deteriorates towards large $q$, where the response of hydrogen even slightly exceeds the response of the full UEG model. This is a direct consequence of the relevant length scales $l=2\pi/q$. Excitations with small $q$ correspond to large distances in coordinate space that only affect free electrons. Large $q$, on the other hand, are directly translated into small $l$, on which bound electrons, too, can react to the external perturbation. 

The blue dots in Fig.~\ref{fig:Response} depict the Kohn-Sham response function $\chi_0(\mathbf{q})$ that has been computed within the local density approximation (LDA)~\cite{Perdew_Zunger_PRB_1981}. For $r_s=2$, these data closely resemble the Lindhard function~\cite{quantum_theory} (dashed black) describing the uniform ideal Fermi gas. This is expected, as the Kohn-Sham orbitals resemble plane waves when localization is weak. The black stars show the DFT response function within the \emph{random phase approximation}, i.e., by setting $G(\mathbf{q},0)=0$ in Eq.~(\ref{eq:chi}). Evidently, this leads to the correct asymptotes for large and small $q$, but becomes inaccurate around $q\sim1.5q_\textnormal{F}$ where electronic XC-effects are known to be important~\cite{review}. 
For $r_s=4$, the Kohn-Sham response function exhibits a more interesting behaviour: for large $q$, it approaches the Lindhard function of the full electronic density, whereas it more closely resembles the reduced Lindhard function corresponding to $\alpha=0.6$ (solid grey) for $q\lesssim2q_\textnormal{F}$. In other words, some information about electron--ion correlations, in general, and about \emph{bound states}, in particular, is included in the KS-response. Remarkably, the corresponding RPA data are in good agreement to the exact PIMC results; more so than for the less strongly coupled case of $r_s=2$.
We also include the KS response function that has been obtained from a DFT simulation without any XC-effects for $r_s=4$ as the grey circles. Interestingly, this leads to substantial differences compared to the blue dots for small to medium $q$, and to a deterioration in the corresponding RPA response function (purple triangles). 

\begin{figure}
\centering
\includegraphics[width=.89\linewidth]{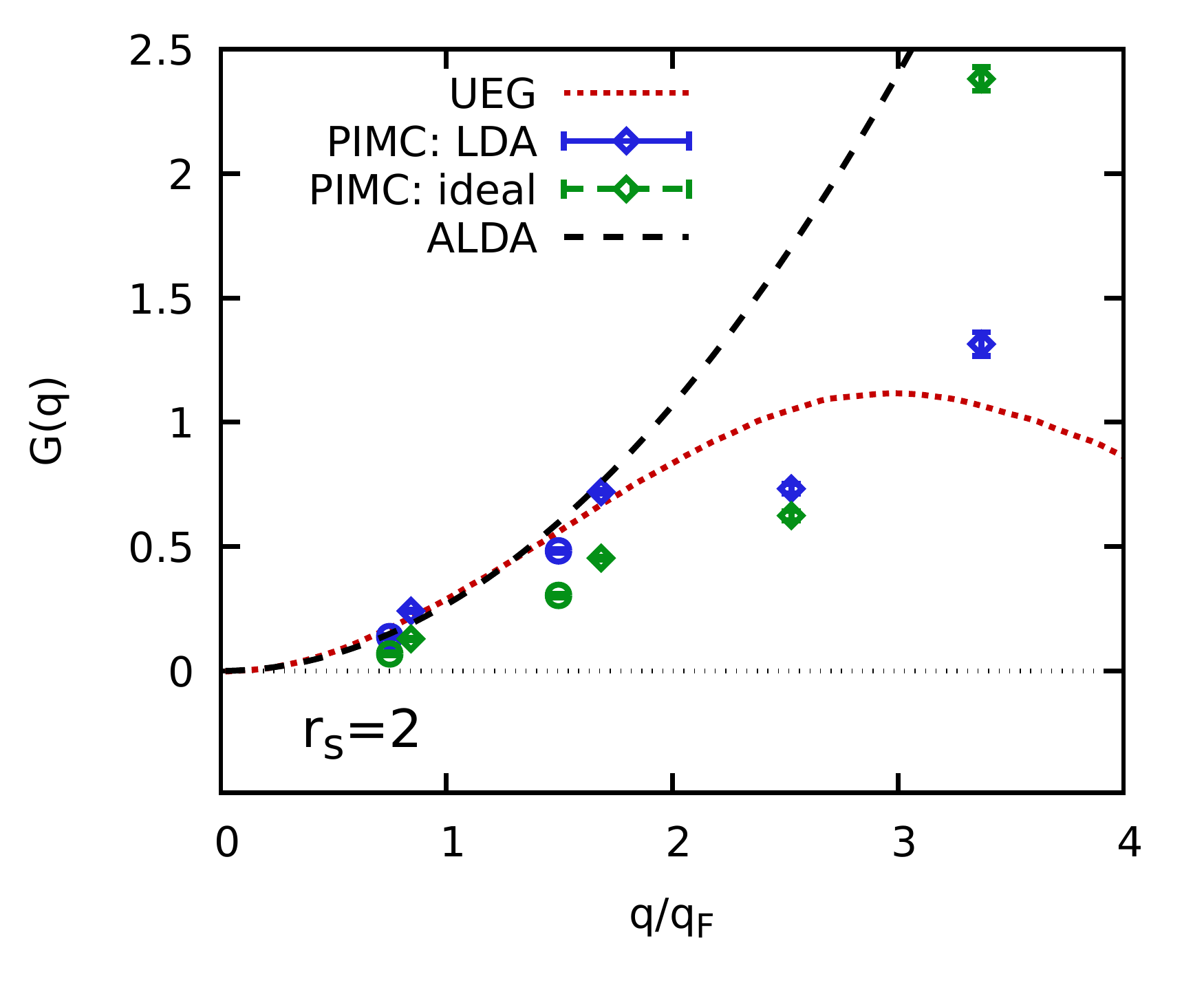}\\\vspace*{-1.17cm}\includegraphics[width=.89\linewidth]{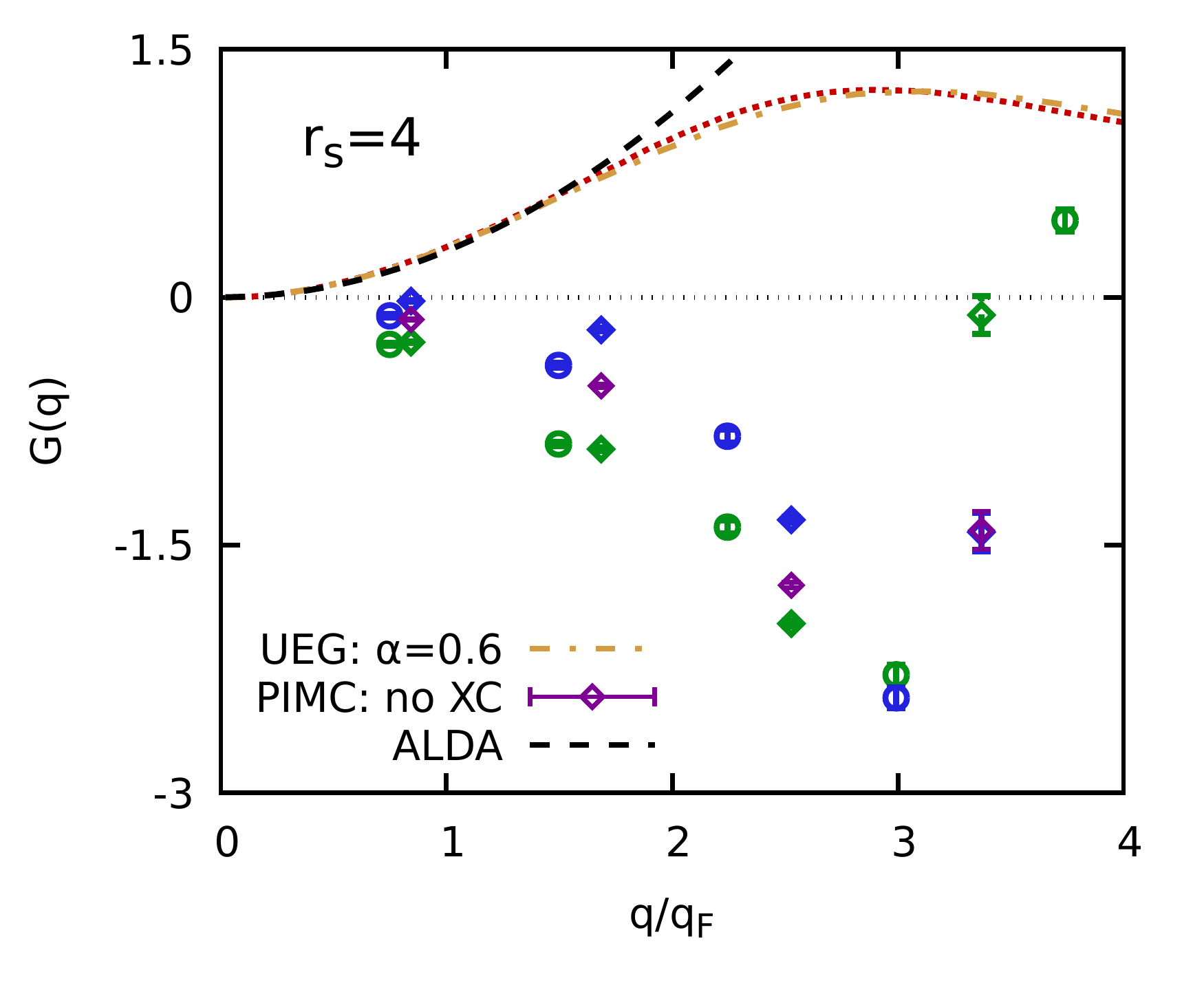}
\caption{Electronic local-field correction $G(q)=G(q,0)$ [cf.~Eq.~(\ref{eq:chi})] of warm dense hydrogen for the same conditions as in Fig.~\ref{fig:Response}. Dotted red lines: UEG model~\cite{dornheim_ML}; blue (green) symbols: solving Eq.~(\ref{eq:chi}) for the LFC using for $\chi_0(\mathbf{q})$ the Kohn-Sham response within LDA (the response function of an ideal uniform Fermi gas~\cite{groth_jcp}). The black dashed line gives the ALDA kernel. Also shown are results for the UEG model under the assumption of a free-electron fraction of $\alpha=0.6$ (dash-dotted yellow) and for the LFC using as input the KS-response computed without any XC effects (purple symbols) for $r_s=4$.
}
\label{fig:LFC}
\end{figure}

Our new PIMC results for the electronic density response of hydrogen give us access to the XC-kernel of a realistic WDM system on a true \emph{ab initio} level. The results are shown in Fig.~\ref{fig:LFC} for the same conditions as in Fig.~\ref{fig:Response}. The dotted red curves show results for the full UEG model~\cite{dornheim_ML} and the dashed black lines the ALDA, which is a parabolic expansion around $q\to0$. 
Evidently, the latter is only appropriate for $q<2q_\textnormal{F}$ even in the case of a UEG. Let us next consider the different results that have been obtained from our PIMC results for hydrogen for $\chi(\mathbf{q})$ by inverting Eq.~(\ref{eq:chi}) using as input different $\chi_0(\mathbf{q})$. The blue (green) symbols have been obtained using the KS response within LDA (the Lindhard function). At $r_s=2$, the DFT kernel agrees well qualitatively with the LFC of the UEG. Interestingly, the kernel that has been computed in terms of the Lindhard function exhibits substantial deviations to the other data sets over the entire $q$-range as it has to balance the absence of electron--ion correlations from $\chi_0(\mathbf{q})$.

For $r_s=4$, the situation is more complicated as none of the data sets computed from $\chi(\mathbf{q})$ resemble the UEG model. For $q\lesssim2q_\textnormal{F}$, the LDA-RPA response function is in good agreement with the PIMC results, and the thus extracted LFC is small in magnitude in this regime. This changes for intermediate $q$, where the LFC becomes comparable in magnitude to the UEG model, but with a negative sign. This is a direct consequence of the overestimation of the actual response by the RPA in the case of hydrogen, whereas it is well known~\cite{dornheim_ML,dynamic_folgepaper} that the opposite holds for the UEG. Using the response function of a uniform ideal Fermi gas to compute the LFC enhances this trend, and the KS-response from a DFT simulation with no XC-effects is located in between. In addition, this means that using either the ALDA or full UEG model as the XC-kernel in a TD-DFT calculation of hydrogen leads to an actually worse estimation of the electronic density response compared to the bare RPA at $r_s=4$.


\textbf{Conclusion:} In this work, we have presented the first, highly accurate PIMC results for the electronic density response of hydrogen in the WDM regime. 
This has allowed us to unambiguously quantify the importance of nonlinear effects, which are even more important compared to the previously investigated case of a uniform electron gas~\cite{Dornheim_PRL_2020,Dornheim_PRR_2021,Dornheim_JCP_ITCF_2021}. Moreover, we have obtained extensive data for the exact linear-response limit of the static density response function $\chi(\mathbf{q},0)$, which have been used to extract the first exact results for the static LFC $G(\mathbf{q})$ and XC-kernel $K_\textnormal{xc}(\mathbf{q})$. We stress that the latter in general depends on the employed $\chi_0(\mathbf{q},\omega)$, which can be either a Lindhard function or a Kohn-Sham response function that depends on the particular choice of the XC-functional.
This has profound consequences in the presence of bound states, where the actual XC-kernel does not even qualitatively resemble the familiar UEG-models. Consequently, the commonly employed ALDA is appropriate when most electrons are free and behave similarly to a UEG, but dramatically breaks down when the degree of electronic localization around the protons is substantial; see the Supplemental Material~\cite{supplement} for a corresponding TD-DFT study.


Our results indicate a strong need for the development and re-evaluation of presently used model XC-kernels such as ALDA. 
A particular advantage of our study is given by the direct possibility to compare our PIMC results for different properties to corresponding DFT results for the same ionic configuration. This, in turn, will give us invaluable lessons regarding the performance of different XC-functionals~\cite{Moldabekov_PRB_2022,Moldabekov_JCP_2021}, and guide the development of new approaches~\cite{kushal,karasiev_importance,groth_prl,ksdt,Karasiev_PRL_2018}. Finally, we re-iterate the capability of our set-up to study \emph{nonlinear effects} in warm dense hydrogen, which may give unprecedented insights into many-body correlation effects in WDM~\cite{Dornheim_JPSJ_2021} and are known to sensitively depend on important system parameters such as the electronic temperature~\cite{moldabekov2021thermal}.

\begin{acknowledgments}

This work was partly funded by the Center for Advanced Systems Understanding (CASUS) which is financed by Germany's Federal Ministry of Education and Research (BMBF) and by the Saxon Ministry for Science, Culture and Tourism (SMWK) with tax funds on the basis of the budget approved by the Saxon State Parliament.
The PIMC calculations were carried out at the Norddeutscher Verbund f\"ur Hoch- und H\"ochstleistungsrechnen (HLRN) under grant shp00026, on a Bull Cluster at the Center for Information Services and High Performance Computing (ZIH) at Technische Universit\"at Dresden, and
on the cluster \emph{hemera} at Helmholtz-Zentrum Dresden-Rossendorf (HZDR).

\end{acknowledgments}

\bibliography{bibliography}

\begin{thebibliography}{74}
\bibitem [{rep()}]{repo}%
  \BibitemOpen
  \href@noop {} {}\bibinfo {note} {A link to a repository containing all PIMC
  raw data will be made available upon publication.}\BibitemShut {Stop}%
\end{thebibliography}
\end{document}


\title{\underline{Supplemental Material:}
Electronic Density Response of Warm Dense Hydrogen:\\ \emph{Ab initio} Path Integral Monte Carlo Simulations
}

\author{Maximilian B\"ohme}
\affiliation{Center for Advanced Systems Understanding (CASUS), D-02826 G\"orlitz, Germany}
\affiliation{Helmholtz-Zentrum Dresden-Rossendorf (HZDR), D-01328 Dresden, Germany}

\affiliation{Technische  Universit\"at  Dresden,  D-01062  Dresden,  Germany}

\author{Zhandos A.~Moldabekov}

\affiliation{Center for Advanced Systems Understanding (CASUS), D-02826 G\"orlitz, Germany}

\affiliation{Helmholtz-Zentrum Dresden-Rossendorf (HZDR), D-01328 Dresden, Germany}

\author{Jan Vorberger}
\affiliation{Helmholtz-Zentrum Dresden-Rossendorf (HZDR), D-01328 Dresden, Germany}

\author{Tobias Dornheim}
\email{t.dornheim@hzdr.de}

\affiliation{Center for Advanced Systems Understanding (CASUS), D-02826 G\"orlitz, Germany}

\affiliation{Helmholtz-Zentrum Dresden-Rossendorf (HZDR), D-01328 Dresden, Germany}

\maketitle

\section*{Path integral Monte Carlo}

\begin{figure}[b!]
\centering
\includegraphics[width=.8\linewidth]{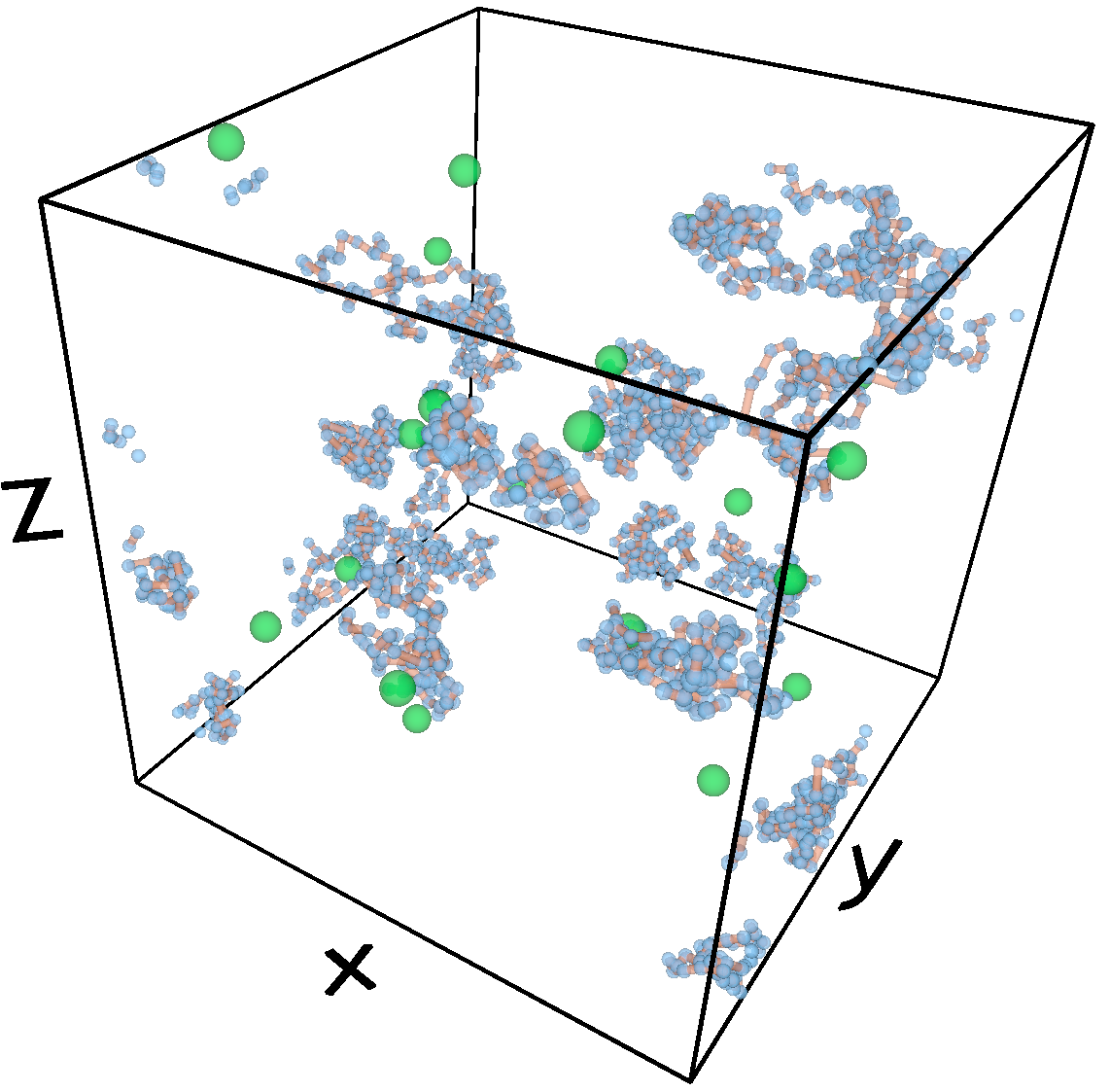}
\caption{Snapshot from a PIMC simulation of hydrogen with $N=20$, $\theta=1$, and $P=200$ for $r_s=2$. The green orbs depict the fixed ions, and the blue-red paths the electronic configuration.}
\label{fig:Snapshot}
\end{figure}

We use the direct PIMC method~\cite{cep,Berne_JCP_1982,Dornheim_permutation_cycles} to simulate the electronic system governed by the Hamiltonian Eq.~(1) of the main text. A typical snapshot is presented in Fig.~\ref{fig:Snapshot}, where we show a configuration of $N=20$ electrons at $\theta=1$ and $r_s=2$. The green orbs represent the (fixed) ions, and the smaller blue-red paths the configuration of the electrons.  We do not impose any nodal restrictions~\cite{Ceperley1991} on the structure of the thermal density matrix. Therefore, our simulations are exact within the given statistical error bars but computationally expensive.

\begin{figure}
\centering
\includegraphics[width=.88\linewidth]{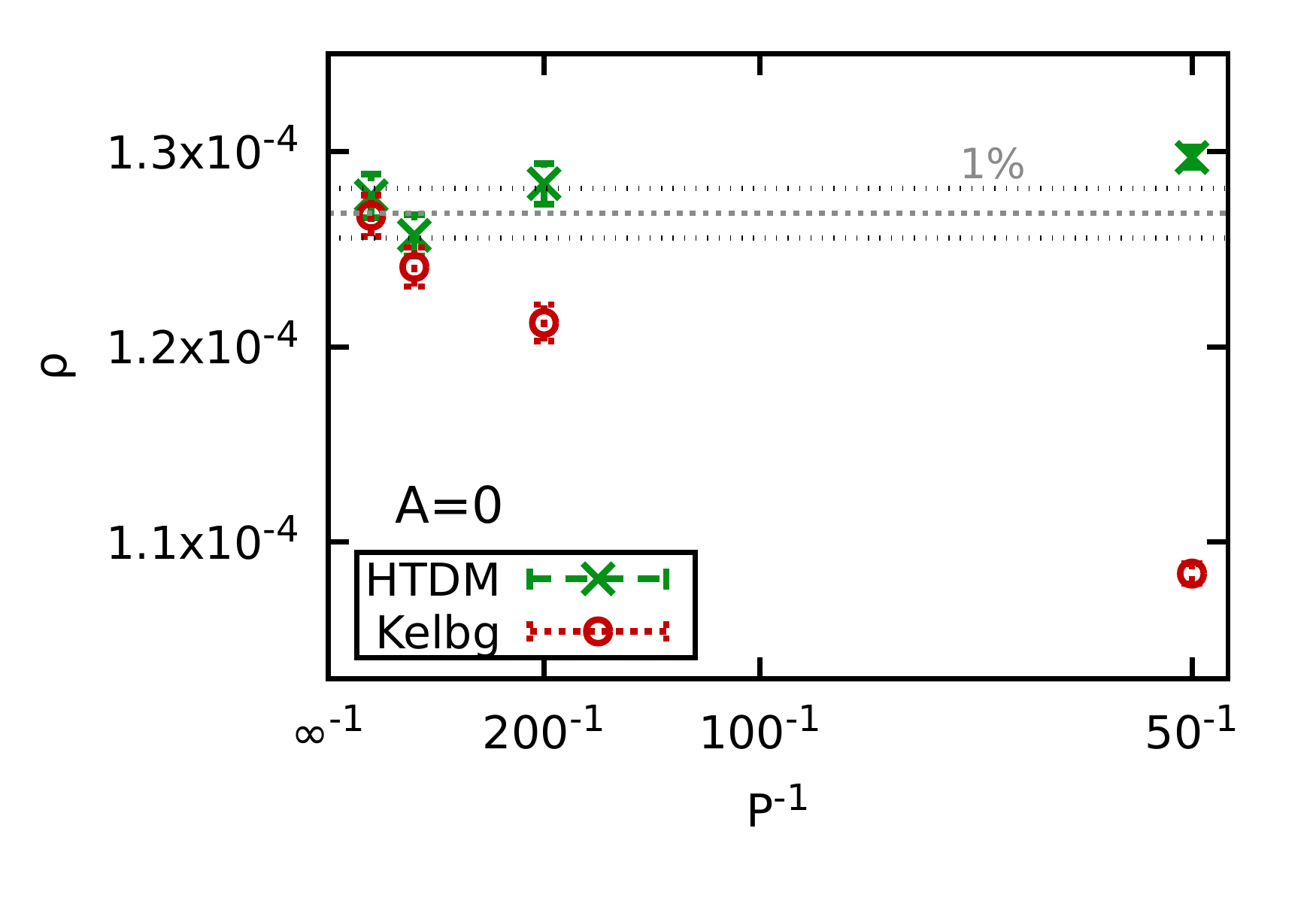}\\\vspace*{-1.2cm}
\hspace*{-0.025\linewidth}\includegraphics[width=.905\linewidth]{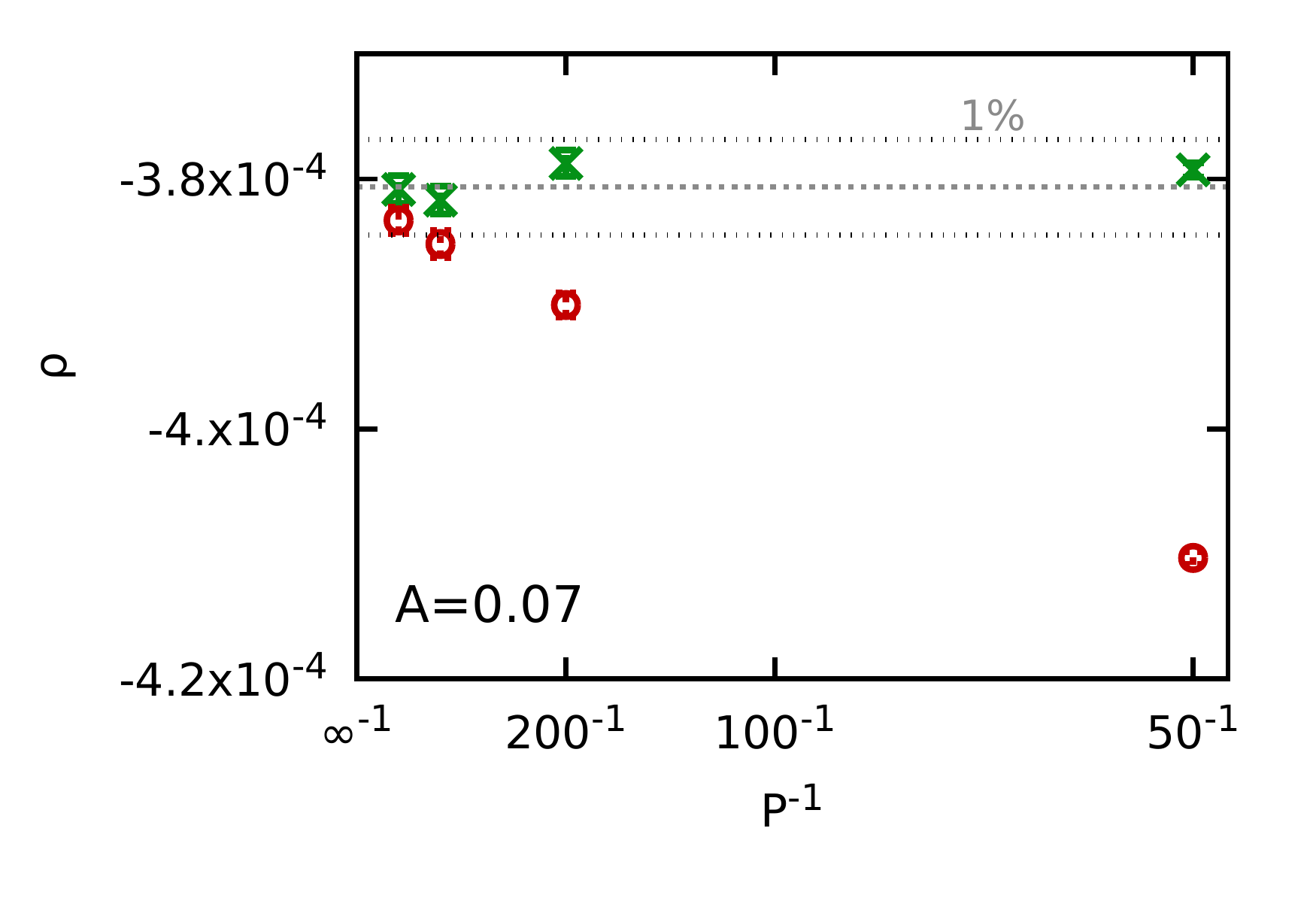}
\caption{
Convergence of the density $\rho(\mathbf{q},A)$ [cf.~Eq.~(\ref{eq:response})] with the number of high-temperature factors $P$ for $N=14$ electrons at $r_s=4$ and $\theta=1$ for $q=3.35q_\textnormal{F}$. Top: $A=0$; bottom: $A=0.07$. The green crosses have been obtained within the pair approximation (used throughout this work), and the red circles correspond to the less efficient Kelbg potential~\cite{Filinov_PRE_2004}.
}
\label{fig:P_convergence}
\end{figure}

Being based on a decomposition of the thermal density matrix, the PIMC method becomes exact only in the limit of an infinite number of imaginary-time steps $P$. In this work, we use the pair approximation that is based on the exact solution of the two-body problem between an electron and an ion following the procedure described by Militzer~\cite{Militzer_HTDM_2016}. 
In Fig.~\ref{fig:P_convergence}, we show the convergence with $P$ of the density in Fourier space $\rho(\mathbf{q},A)$ [see the main text] at $r_s=4$ and $q=3.35q_\textnormal{F}$. The top and bottom panels show results for the unperturbed case ($A=0$), and a perturbation of $A=0.07$. The green crosses correspond to our implementation of the pair approximation, and the propagator error is small even for as few as $P=50$. To ensure that any residual convergence error is negligible, we use $P=500$ throughout this work. As an additional benchmark, we have also implemented a decomposition of the density matrix based on the diagonal Kelbg potential~\cite{Filinov_PRE_2004}. The results are shown as the red circles in Fig.~\ref{fig:P_convergence} and exhibit a substantially larger propagator error compared to the pair approximation, as it is expected. At the same time, they nicely converge towards the green crosses in the limit of $P\to\infty$, which is a strong indication for the correctness of our implementation.

\section*{Impact of the PIMC XC-kernel on TD-DFT simulations}

\begin{figure}[t]
\centering
\includegraphics[width=.79\linewidth]{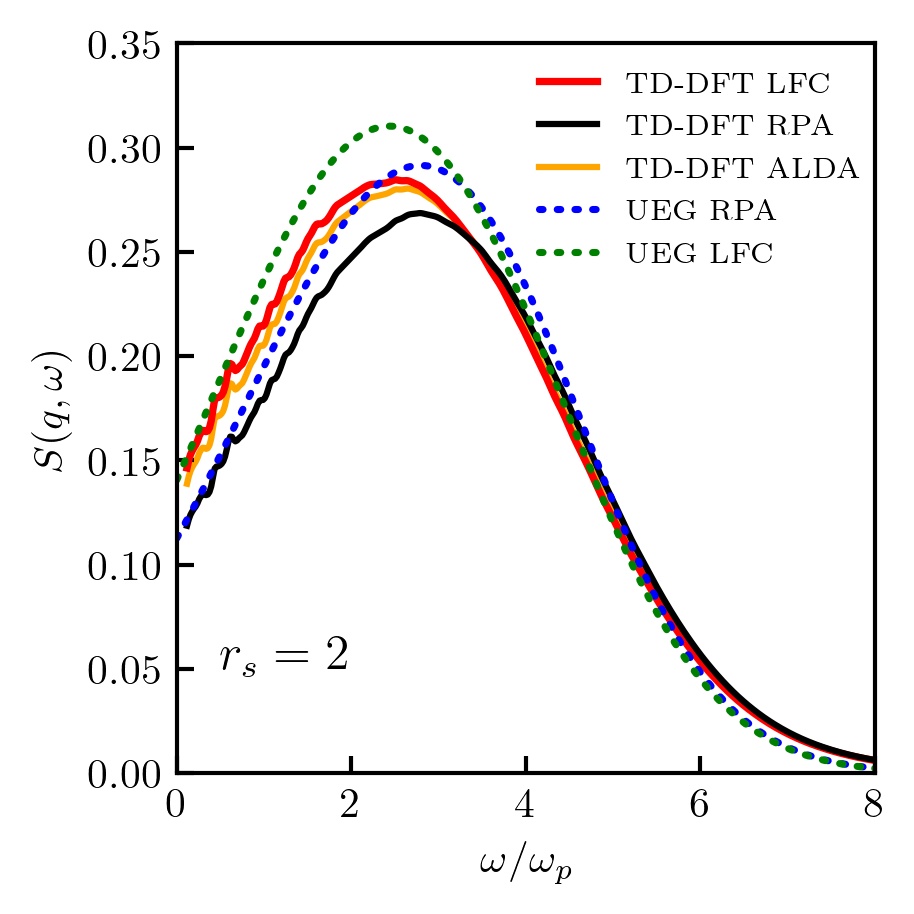}\\\includegraphics[width=.79\linewidth]{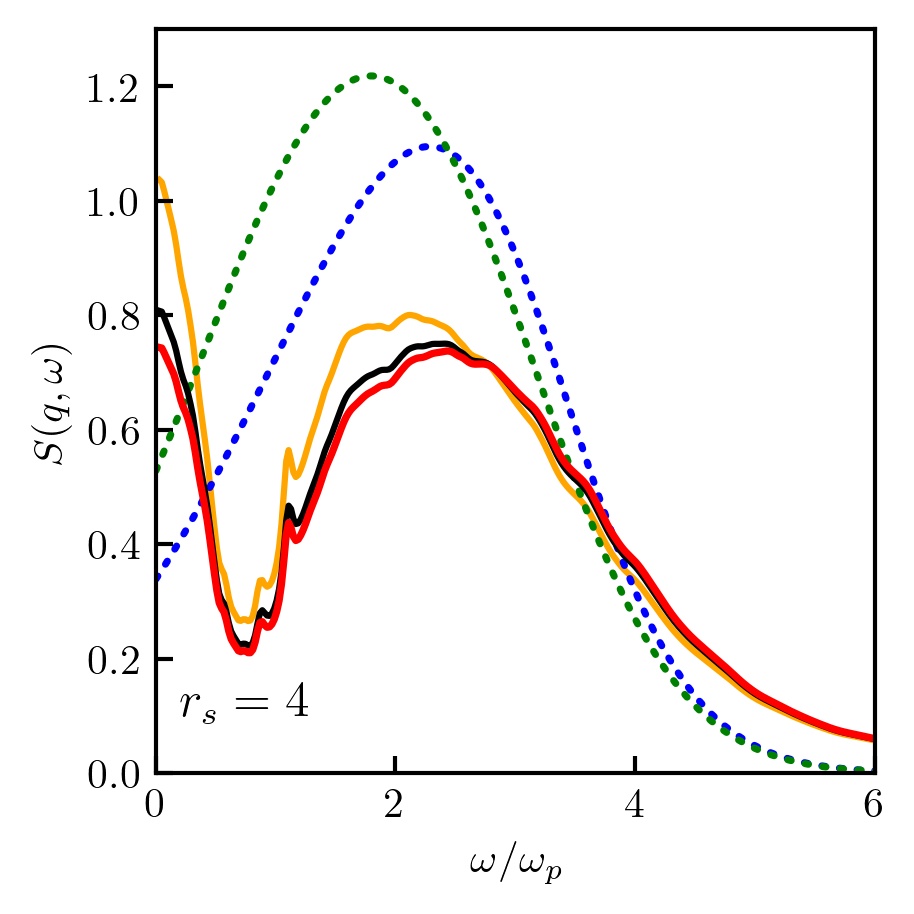}
\caption{TD-DFT results for the dynamic structure factor $S(\mathbf{q},\omega)$ at the electronic Fermi temperature ($\theta=1$) at $q\approx1.69q_\textnormal{F}$ (i.e. $q\approx 3.06 ~\textnormal{\AA}^{-1}$ at $r_s=2$ and $q\approx 1.52 ~\textnormal{\AA}^{-1}$ at $r_s=4$) for $r_s=2$ with $T\simeq 12.5 ~{\rm eV}$ (top) and $r_s=4$ with $T\simeq 3.1 ~{\rm eV}$ (bottom). Dotted blue (dotted green): uniform electron gas without (with) LFC taken from Ref.~\cite{dornheim_ML}. Solid black: TD-DFT RPA; solid red: TD-DFT with consistent PIMC kernel; solid yellow: ALDA. The electronic plasma frequency is given by $\omega_\textnormal{p}=\sqrt{3/r_s^3}$.
}
\label{fig:TDDFT}
\end{figure}

We have carried out linear-response time-dependent DFT (TD-DFT) calculations to assess the impact of the electronic XC-kernel on the dynamic structure factor (DSF) $S(\mathbf{q},\omega)$ of the electrons in hydrogen; all relevant DFT parameters are listed below.
The results are shown in Fig.~\ref{fig:TDDFT}, and the top (bottom) panel shows the DSF at $\theta=1$ for $q\approx1.69q_\textnormal{F}$ at $r_s=2$ ($r_s=4$). The dotted blue and green curves depict corresponding results for the UEG~\cite{dornheim_dynamic,dynamic_folgepaper} that have been obtained within RPA, and by using the static LFC of the UEG, respectively. Evidently, the incorporation of the latter leads to a correlation induced red-shift in both cases, which is more pronounced at  the lower density, as it is expected. The $\omega$-dependence of $G(\mathbf{q},\omega)$ has a negligible impact on the DSF of the UEG at these conditions, and the dotted red curves are exact for a pure electron gas without the ions~\cite{dornheim_dynamic,dynamic_folgepaper}. 
The solid black and red curves show TD-DFT results (averaged over $N_\textnormal{snap}=13$ independent snapshots) based on LDA calculations on the RPA level and using the appropriate exact static XC-kernel based on PIMC results for hydrogen shown in Fig.~3 in the main text, respectively. To explicitly focus on the impact of $K_\textnormal{xc}(q)$, we compute the electron--electron part of the full dynamic structure factor of hydrogen, which, however, intrinsically depends on the ionic structure. The full DSF includes an additional electron--ion term (leading to the well-known inelastic peak) which we do not consider in this work. For $r_s=2$, both curves are qualitatively similar to the corresponding UEG curves, as it is by now expected. In stark contrast, we observe substantial qualitative differences to the UEG at $r_s=4$. In particular, our TD-DFT results for $S(\mathbf{q},\omega)$ exhibit a  diffusive peak induced by the localization of the electrons in the attractive potential of the ions
around $\omega=0$ (see also Refs.~\cite{Choi_PRE_2021,Dornheim_SciRep_2022} for observations of such a feature in other systems) and a second peak around twice the electronic plasma frequency, $\omega=2\omega_\textnormal{p}$. Interestingly, the incorporation of the appropriate electronic LFC has a negligible impact onto the TD-DFT results, as the RPA is nearly exact in the static limit, see the main text. We re-iterate our earlier observation that this is the opposite trend compared to the UEG, where the impact of the LFC increases monotonically with $r_s$. 
Finally, the yellow curves show TD-DFT results for $S(\mathbf{q},\omega)$ that have been obtained on the basis of the ALDA model. Evidently, this approximation is well justified for $r_s=2$, where hydrogen behaves similarly to an UEG. Contrary, ALDA leads to a significant deterioration of the quality of $S(\mathbf{q},\omega)$ at $r_s=4$. In order words, it can indeed be preferable to neglect the XC-kernel altogether, instead of employing models that have been derived in a different context, and under assumptions that can be strongly violated in realistic WDM applications.

\section*{Density functional theory parameters}

The KS density response function
is computed using the KS eigen wavefunctions  $\psi_{n \mathbf{k}}$ with 
eigenvalues $\epsilon_{n \mathbf{k}}$ as~\cite{PhysRev.126.413, PhysRev.129.62}:
\begin{eqnarray}\label{eq:chi0}
 \chi^0(\mathbf{q}, \omega) &=& \frac{1}{\Omega}
\sum_{\mathbf{k}, n, n^{\prime}}
\frac{f_{n\mathbf{k}}-f_{n^{\prime} \mathbf{k} + \mathbf{q} }}{\omega + \epsilon_{n\mathbf{k}} - \epsilon_{n^{\prime} \mathbf{k} + \mathbf{q} } + i\eta}\\
& & \times
 \langle \psi_{n \mathbf{k}} | e^{-i\mathbf{q} \cdot \mathbf{r}} | \psi_{n^{\prime} \mathbf{k} + \mathbf{q} } \rangle
 \langle \psi_{n\mathbf{k}} | e^{i\mathbf{q} \cdot \mathbf{r}^{\prime}} | \psi_{n^{\prime} \mathbf{k} + \mathbf{q} } \rangle. \nonumber
\end{eqnarray}

The KS-DFT calculations have been performed using the GPAW code~\cite{GPAW1, GPAW2, ase-paper, ase-paper2}, which is a real-space implementation of the projector augmented-wave method.  
A Monkhorst-Pack~\cite{PhysRevB.13.5188} sampling of the Brillouin zone was used. The calculations are carried out using a plane-wave basis. 

For the calculation of the \textit{static density response function}  $\chi^0(\mathbf{q}, 0)$, the following parameters have been used:

For $r_s=2$, $\theta=1$ ($T=12.5280 ~{\rm eV}$), and the number of particles  $N=14$ ($N=20$), the main simulation cell size is $L= 4.1 ~\AA$ ($L= 4.631 ~\AA$) with a spacing of $h=0.1708~ \AA$ ($h=0.193~ \AA$), the cut-off energy in the thermal equilibrium calculation  is set to be $E_{\rm cut}= 550~{\rm eV}$ ($E_{\rm cut}= 650~{\rm eV}$),  the number of bands is given by $1500$ ($1100$),  and a \emph{k}-point grid of $4\times4\times4$. 
The wavefunctions from the thermal equilibrium calculation are used to compute the static density response function $\chi^0(\mathbf{q}, 0)$ using a plane-wave cut-off of  $325~{\rm eV}$ ($150~{\rm eV}$),  
where the wave numbers $q=jq_{\rm min}$, with $j=1...5$ ($j=1...4$) and $q_{\rm min}=2\pi/L$,  have been considered.
The data point for $q_{\rm min}/12$ is generated using  $300$ bands, with $E_{\rm cut}= 550~{\rm eV}$, $h=0.185 \AA$, and  a $12\times12\times12$ \emph{k}-point grid.

For $r_s=4$, $\theta=1$ ($T=3.132 ~{\rm eV}$), and the number of particles $N=14$ ($N=20$),
the main simulation cell size is $L= 8.224~ \AA$ ($L= 9.262~ \AA$) with a spacing $h=0.25 \AA$, the cut-off energy in the thermal equilibrium calculation is set to be $E_{\rm cut}= 400~{\rm eV}$,  the number of bands $1900$ ($1300$),  and the \emph{k}-point grid $4\times4\times4$. 
$\chi^0(\mathbf{q}, 0)$ is computed using a plane-wave cut-off of $90~{\rm eV}$,  
where the wave numbers $q=jq_{\rm min}$, with $j=1...5$ ($j=1...4$),  have been considered.
The data point for $q_{\rm min}/16$ is generated using  $250$ bands, $E_{\rm cut}= 350~{\rm eV}$, $h=0.25 \AA$, and  a $16\times16\times16$ \emph{k}-point grid.

For the calculation of the \textit{dynamic density response function}  $\chi^0(\mathbf{q}, \omega)$ that has been used to compute $S(q,\omega)$ shown in Fig.~\ref{fig:TDDFT}, the following parameters are used:

At $r_s=2$ ($r_s=4$), $\chi^0(\mathbf{q}, \omega)$ is computed using  the main simulation cell size $L= 4.1 ~\AA$ ($8.224~ \AA$ ) with a spacing of $h=0.25~ \AA$, the cut-off energy in the calculation of the equilibrium state is set to be $E_{\rm cut}= 350~{\rm eV}$,  the number of bands $500$ ($400$),  and the \emph{k}-point grid of $12\times12\times12$ ($4\times4\times4$). 
The wavefunctions from the equilibrium state calculation are used to compute $\chi^0(\mathbf{q}, \omega)$ with a plane-wave cut-off of $80~{\rm eV}$ and the broadening parameter $\gamma=0.2~{\rm eV}$.



\section*{References}
\bibliography{bibliography}